\input phyzzx
\PHYSREV
 
\def\IR {{\rm I\!R}}


\def\a {\alpha}
\def\b {\beta}
\def\g {\gamma}
\def\G {\Gamma}
\def\O {\Omega}
\def\d {\delta}
\def\e {\epsilon}
\def\f {\psi}

\def\l {\lambda}

\def\S {\Sigma} 

\def\t {\tau}
\def\te {\theta}
\def\vf {\chi}


\def\dif {\rm d}
\def\w {\wedge}
\def\half {\coeff{1}{2}}
\def\vect #1{\overrightarrow {#1}}
\def\sh {\sinh}

\def\twid #1{\tilde {#1}}

\def\dg {\dot g}
\def\ba {{\b} ^0}
\def\bb {{\b} ^+} 
\def\bc {{\b} ^-} 
\def\db {\dot {\b}}
\def\dba {\dot {\ba}}
\def\dbb {\dot {\bb}} 
\def\dbc {\dot {\bc}} 
\def\du {\dot u}
\def\dv {\dot v}
\def\dte {\dot {\te}}
\def\pa {p _0}
\def\pb {p _+}
\def\pc {p _-}

\def\H {\it H}
\def\N {\it N}
\def\nen {\vect {\N}}
\def\Md {\dot {M _E}}

\def\Lt {\twid L}
\def\x {\xi}
\def\y {\eta}
\def\pde #1#2{\partial_{#2} #1}
\def\z {\phantom z}

\REF\G{R.H. Gowdy, Ann. Phys. 83 (1974) 203 }

\REF\VER{E. Verdaguer, Phys. Rep. 229 (1993)}

\REF\GER{R. Geroch, J. Math. Phys. { 12} (1971) 918;
{ 13} (1972) 394}

\REF\HB{B. K. Harrison, Phys. Rev. Lett. {41} (1978) 1197;
Phys. Rev. D { 21} (1980) 1695}

\REF\BZ{Belinskii and Zakharov, Sov. Phys. JETP 48 (1978) 985}

\REF\DM{D. Maison, Phys. Rev. Lett. 41 (1978) 521}

\REF\HE{J. Hauser and F. J. Ernst, Phys. Rev. D { 20} (1979) 362;
J. Math. Phys. { 21} (1980) 1126; { 22} (1981) 1051}

\REF\MS{N. Manojlovi\'c and B. Spence, Nucl. Phys. B 423 (1994) 243}

\REF\KN{D. Korotkin and H. Nicolai, Nucl. Phys. B 475 (1996) 397}

\REF\KS{D. Korotkin and H. Semtleben, Phys. Rev. Lett. 80 (1998)14 }

\REF\CMN{J. Cruz, A. Mikovi\'c and J. Navarro-Salas, Phys. Lett. B 437 (1998) 
273}

\REF\BF{V. Belinskii and M. Francaviglia, Gen. Rel. and Grav. 14 (1982) 213 }

\REF\AS{A. Ashtekar and J. Samuel, Class. Quant. Grav. 8 (1991) 2129}

\REF\JA{Y. Fujiwara, H. Kodama and H. Ishihara, Class. Quant. Grav. 19 (1993) 859}

\REF\J{T. Koike, M. Tanimoto, A. Hosoya, J. Math. Phys. 35 (1994) 4855 }

\REF\K{H. Kodama, Prog. Theor. Phys. 99 (1998) 173}

\REF\JAN{R.T. Janzen, Commun. Math. Phys. 64 (1979) 211 }

\REF\MM{N. Manojlovi\'c and A. Mikovi\'c, Nucl. Phys. B 382 (1992) 148}

\REF\FN{H. Flaschka and A.C. Newell, Commun. Math. Phys. 76 (1980) 65}

\REF\JMU{M. Jimbo, T. Miwa and K. Ueno, Physica D 2 (1981) 306}

\REF\I{A.R. Its and V.Y. Novokshenov, The Isomonodromic Deformation Method in
the Theory of Painlev\'e Equations, Lecture Notes in Mathematics, 
Springer-Verlag, Berlin (1986)}

\REF\IN{E.L. Ince, Ordinary Differential Equations, Dover Publications, 
New York (1956)}

\date={August 1999}

\titlepage

\title{\bf Painlev\'e III Equation and Bianchi VII$_0$ Model}

\author{Nenad Manojlovi\'c\foot{E-mail:nmanoj@ualg.pt} and 
Aleksandar Mikovi\'c\foot{On leave 
of absence from Institute of Physics, Belgrade, Yugoslavia }
\foot{E-mail:amikovic@ualg.pt}}
\address{\'Area Departamental de Matem\'atica, UCEH, Universidade do Algarve, Campus de 
Gambelas, 8000 Faro, Portugal}

\abstract{We examine the reduced phase space of the Bianchi VII$_0$ 
cosmological model, including the moduli sector.
We show that the dynamics of the relevant sector of local degrees of freedom 
is given by a Painlev\'e III equation. We then obtain a zero-curvature 
representation of this  Painlev\'e III equation by applying
the Belinskii-Zakharov method to the Bianchi VII$_0$ model.}
\endpage

{\bf \chapter {Introduction}}

Bianchi VII$_0$ cosmological model is a finite-dimensional dynamical
system with non-trivial dynamics. It can be considered 
as a symmetry reduction of the Gowdy model [\G]
whose spatial hypersurface is the three torus $T^3$. The Gowdy models 
represent 
examples of two commuting Killing reductions of general relativity [\VER], 
whose
space-times have compact spatial hypersurfaces. Although
the two commuting Killing vector system is integrable 
[\GER, \HB,\BZ,\DM,\HE], it is not known how to extend the corresponding 
methods of constructing solutions [\MS,\KN,\KS,\CMN] to the compact case 
topology. Therefore Bianchi VII$_0$ model can serve
as a toy model for exploring the issue of integrability of the Gowdy model
on the three torus. 

Belinskii and Francaviglia showed, within a more general framework of 
Belinskii and Zakharov inverse scattering method  [\BZ], that 
the Einstein equations for some Bianchi models admit a zero-curvature 
representation, indicating that these are solvable dynamical systems [\BF]. 
However, their analysis was not complete in two aspects.
The first aspect is related to the fact that their considerations were only 
local, since
they ignored topological obstructions coming from the non-trivial global 
topology
of the spatial hypersurface, the three torus T$^3$. The problem is that spacetimes 
with compact spatial sections do not allow in general global Bianchi metrics, 
and one can put only locally homogeneous metrics [\AS,\JA]. Related to that is 
that compact spatial manifolds have non-trivial 
topology, and locally
diffeomorphic metrics are not necessarily globally diffeomorphic, which means that 
there are global
degrees of freedom in the metric, beside the usual local ones. 
It has been shown recently that the moduli parameters
enter non-trivially in the diffeomorphism invariant symplectic form, and hence they 
could change the dynamics of the local degrees of freedom [\J,\K].
The second unexplored aspect is to find out what kind of integrable nonlinear
dynamical equation can be obtained.   

In order to explore these issues, it is convenient
to study the dynamics of Bianchi VII$_0$ model in the canonical formalism.
We perform constraint and 
gauge-fixing analysis and show that 
the dynamics of a generic sector of local degrees of freedom can 
be  reduced to that of a Painlev\'e III equation. There is also a special 
sector with
enhanced symmetry, which has a linear dynamical equation.
By using the results of Kodama [\K] we show that  
the moduli parameters do not change the dynamics of generic local 
degrees of freedom.
We then show how Bianchi VII$_0$ model appears in the Belinskii-Zakharov 
approach, and 
how to obtain Painlev\'e III equation. We then  use these results to obtain 
a linear system whose zero-curvature condition is a Painlev\'e III equation.

{\bf \chapter {The Class A Bianchi Models}}

Bianchi models are spatially homogeneous spacetimes which admit
a three dimensional isometry Lie group G that acts simply transitively
on each leaf ${\S}$ of the homogeneous foliation,
for a review and references see [\JAN]. As a consequence,
there exists for each of these models a set of three left-invariant
vector fields $L _I$ on ${\S}$ which form the Lie algebra of the
group $G$:
$$\eqalignno {
[ L _I , L _J ] &= {C ^K} _{IJ} L _K  \quad ,  &\eqnalign\liealgebra \cr}$$
where ${C ^I} _{JK}$ are the structure constants of the Lie group.

Dual to the the vector fields $L _I$, one can introduce a set of
three left-invariant one-forms $\vf ^I$ which satisfy the
Maurer-Cartan equations
$$\eqalignno {
{\dif} {\vf} ^I + \half \, {C^I} _{JK} \, \, \, \vf ^J \w \vf ^K &= 0
					\quad .  &\eqnalign\mce \cr}$$

If the trace ${C ^I} _{IJ}$ of the structure constants is equal to
zero, the Bianchi model is said to belong to Bianchi class {\bf A}.
For this class of models, the spacetime admits foliations by
compact slices. 

	The structure constants for the class {\bf A} Bianchi models
can always be written in the form
$$\eqalignno {
{C ^I} _{JK} &= \e _{JKL} S ^{LI}  \quad , &\eqnalign\stconst \cr}$$
where $\e _{JKL}$ is the totally antisymmetric symbol, and $S ^{IL}$ is a 
symmetric
tensor density of weight one over the Lie algebra of $G$. Further
classification of the class {\bf A} Bianchi models
is defined with respect to the signature of the symmetric 
tensor density $S^{IJ}$.
The type VII Bianchi model of class {\bf A} is denoted as Bianchi VII$_0$, and
it is characterized by the signature
$( + , + , 0 )$. Hence the structure constants for this model
are given by
$$
\eqalignno {
{C ^I} _{JK} &= \e _{JK1} \, \d ^I_1 + \e _{JK2}\, \d ^I_2 
				\quad . &\eqnalign\stconstbvii \cr}
$$

Bianchi models 
can be considered as the homogeneous sector of general relativity.
The dynamics can be obtained by performing the corresponding reduction 
of the canonical formulation 
of general relativity. The canonical variables of general relativity are the three 
metric
$g _{ij} (t , x ^i )$ on the spatial section $\Sigma$, and its canonically conjugate 
momenta 
${\pi} ^{ij} (t , x ^i )$,
where $x^i$ are coordinates on  $\Sigma$. The action for a
Bianchi model can be obtained by inserting the expressions
$$
\eqalignno { g _{ij} (t,x) &= g_{IJ}(t) \chi^I_i (x) \chi^J_j (x)  \quad , \quad
{\pi}^{ij} (t,x)= \pi^{IJ}(t) L_I^i (x) L_J^j (x)  
			\quad  &\eqnalign\homoggp \cr}$$
into the canonical form of the Einstein-Hilbert action. This gives
$$
\eqalignno { S _{BA} &= \int _{t_0}^t {\dif} t \, \Bigl ( {\pi}  ^{IJ}
{\dot g}_{IJ} - {\N} ^I {\H} _I - {\N} \, {\H} _0  \Bigr ) 
			\quad , &\eqnalign\baaction \cr}
$$
where the
canonical variables $( g_{IJ} \, , \, {\pi}^{IJ} )$ have the
Poisson brackets
$$
\eqalignno { \{ \, \, g_{IJ} \, , \, {\pi}  ^{KL} \} &= {\half} \,     
\bigl ( \, {\d} _I^K \, {\d} _J^L + {\d} _I^L \, {\d} _J^K \bigr)
				\, . &\eqnalign\fundpb \cr}
$$
The vector constraint ${\H}_I$, I = 1, 2, 3, is the reduction of the
diffeomorphism constraint, and $H_I$ is given by
$$
\eqalignno { {\H} _I &= 2 \, {C ^J} _{KI} \, g_{JL} \, {\pi}  ^{LK}
= 2 \, \e _{KIM} S ^{MJ} \, g_{JL} \, {\pi}  ^{LK}	\approx 0     
					\, . &\eqnalign\badiff \cr}
$$
The reduction of the Hamiltonian constraint is ${\H}_0$, and it is given by
$$
\eqalignno { {\H} _0 &= {1\over{\sqrt {\det g}}} \Bigl ( 
{\Tr} (  g {\pi} g {\pi} ) 
- {\half} {\Tr} ^2 ( g {\pi}) + {\Tr} ( g S gS )
- {\half} {\Tr} ^2 ( g S ) \Bigr ) \approx 0 
			\, .  &\eqnalign\baham \cr}
$$
${\N} ^I$ and ${\N}$ are the corresponding Lagrange multipliers.
The constraints {\badiff} and {\baham}  form a closed Poisson algebra
$$
\eqalignno { \{ {\H} _I \, , \, {\H} _J \} &= {C ^K} _{IJ} \, {\H} _K   
			\quad , &\eqnalign\bapalgd \cr
\{ {\H} _0 \, , \, {\H} _I \} &= 0	
			\quad , &\eqnalign\bapalghd \cr}
$$
and therefore they constitute a set of first-class constraints. 
Furthermore, the equation {\bapalghd} implies that the
hamiltonian constraint is invariant under the transformations 
generated by the vector constraint. This fact can be used to
go to the parametrized particle form of the action,
defined by the diffeomorphism invariant variables and the 
Hamiltonian constraint.

{\bf \chapter {Vector Constraint}}

The standard approach to finding the dynamics of Bianchi models is to solve
first the vector constraint. This requires the corresponding gauge-fixing, and
because of {\bapalghd}, one can even find the diffeomorphism invariant 
variables explicitly.
  
In the case of Bianchi VII$_0$ model the topology of $\S$ is fixed to be
the three torus $T ^3$ and the coordinates $x^i = ( x , y , z )$
can be chosen such that ${\vf} ^1 , {\vf} ^2, {\vf} ^3$ have the
canonical form
$$
\eqalignno {   {\vf} ^1 &= \cos z  \, {\dif} x  + \sin z \, {\dif} y
                                           \, , \quad 
               {\vf} ^2 = - \sin z \, {\dif} x + \cos z  \, {\dif} y
                                           \, , \qquad
               {\vf} ^3 = {\dif} z
			 \quad .  &\eqnalign\formstorus \cr}
$$
The three left-invariant one-forms $\vf ^I$ satisfy the
Maurer-Cartan equations for the type VII Bianchi model

$$
\eqalignno {
{\dif} {\vf} ^1 + \, \vf ^2 \w \vf ^3 &= 0 \, , \quad
{\dif} {\vf} ^2 + \, \vf ^3 \w \vf ^1 = 0  \, , \quad
{\dif} {\vf} ^3 = 0		\, .  &\eqnalign\formsvii \cr}
$$

Note that every Bianchi model has a symmetry group $M$, which is the automorphism group
of the Lie algebra of $G$. $M$ can can be realized as
a subgroup of ${\bf {GL}} ( 3 \, , {\IR})$ in the following way.
Let us consider a set of the left-invariant vector fields $L_I$. They form the Lie 
algebra of $G$ through the commutation
relations {\liealgebra}. An invertible matrix 
${M ^I} _J$ yields a new set of left-invariant vector fields
${\Lt}_J = L_I M^I_J$, whose commutation relations will have  the same 
structure constants as $L_I$
if the following identity is satisfied
$$
\eqalignno { {C ^I} _{JK} &= {(M ^{-1}) ^I} _L  \, {C ^L} _{MN} \, 
{M ^M} _J \, {M ^N} _K 		\quad , &\eqnalign\stconstsym \cr}
$$
where ${(M ^{-1}) ^I} _L$ is the inverse of the matrix ${M ^L} _I$. The matrices
$M^I_J$ with the condition
{\stconstsym} define the symmetry group $M$.

Equivalently, with the help of the identity {\stconst}, we can define $M$ as
$$
\eqalignno { S ^{IJ} &= \bigl ( {\det} M \bigr ) ^{-1} \,
{M ^I} _K \, S ^{KL} \, {\bigl ( M ^T \bigr ) _L} ^J 
                                    \quad , &\eqnalign\ssym \cr}
$$
where ${(M ^T) _L} ^J$ is the transpose of the matrix ${M ^J} _L$. 
In particular, in the case of the Bianchi-VII model the
tensor density $S ^{IJ}$ has the signature $( + , + , 0 )$  
(see equation {\stconstbvii}), and hence the condition {\ssym} 
implies the following form of the matrix $M$
$$
\eqalignno { M &= M _D \, M _E = 
\left ( \matrix { e ^{{3\over 2} c_0} & 0 & 0           \cr
                  0 & \pm e ^{{3\over 2} c_0} & 0           \cr        
                            0 & 0 & \pm 1                   \cr} \right )
\, 
\left ( \matrix { \cos {\te} & \sin {\te} & u           \cr
               - \sin {\te} & \cos {\te} & v            \cr        
                            0 & 0 & 1                   \cr} \right )
                                         \quad . &\eqnalign\mmatrix \cr}
$$

Let us consider the following change of variables
$$
\eqalignno { g_{IJ} ( t ) &= {\bigl ( {M_E} ^T \bigr ) _I} ^K ( t ) \, 
Q_{KL} ( t ) \, {\bigl (M_E \bigr ) ^L} _J ( t )
					\quad , &\eqnalign\newg \cr}
$$
where ${(M_E ) ^I} _J ( t )$ is given by 
$$
\eqalignno { M _E ( t ) &= 
\left ( \matrix { \cos {\te} (t) & \sin {\te} (t) & u (t)   \cr
                - \sin {\te} (t) & \cos {\te} (t) & v (t)   \cr   
                            0 & 0 & 1                       \cr} \right )
                                   \quad , &\eqnalign\newm \cr}
$$
and $Q_{IJ}={\rm diag}(Q_1, Q_2, Q_3)$ is a diagonal matrix. It is useful to 
introduce
new variables $(\b_0 ,\b_+ ,\b_- )$ as  
$$
\eqalignno { \!\!\!\! Q_{1} &= 
e ^{2 \, ({\ba} + {\bb} + {\sqrt 3} \, {\bc})} \, ,  \, \,     
Q_{2}  = e ^{2 \, ({\ba} + {\bb} - {\sqrt 3} \, {\bc})} 
\, , \, \, 
Q_{3} = e ^{2 \, ({\ba} - 2 {\bb})}  
                        \, . &\eqnalign\gtcomp \cr}
$$
The definition {\newg} involves only $M_E$, because by rescaling of $Q$ by constants we 
can always put $M_D = Id$.

We have two alternative ways of imposing the constraints ${\H} _I$
{\badiff}. One way is to complete the canonical 
transformation {\newg} by calculating the conjugate momenta
$$
\eqalignno { p _{\te} &= \partder {g _{IJ}} {\te} \, {\pi} ^{IJ}
\quad , \quad p _u =\partder {g _{IJ}} {u} \, {\pi} ^{IJ} \quad , 
\quad p _v = \partder {g _{IJ}} {v} \, {\pi} ^{IJ}
				\quad , &\eqnalign\momtuv \cr
p _0 &= \partder {g _{IJ}} {\ba} \, {\pi} ^{IJ}   \quad , \quad 
p _+ = \partder {g _{IJ}} {\bb} \, {\pi} ^{IJ}    \quad , \quad
p _- = \partder {g _{IJ}} {\bc} \, {\pi} ^{IJ}
				\quad , &\eqnalign\mombabc \cr}
$$
and then expressing the constraints {\badiff} in terms of 
the new canonical pairs $( {\te} \, , \, p _{\te} \, , \,  
u \, , \, p _u \, , \,  v \, , \, p _v \, , \,  {\ba} \, , \, {\pa} 
\, , \,  {\bb} \, , \, {\pb} \, , \,  {\bc} \, , \, {\pc} )$. This is the approach
taken by Kodama [\K]. Alternatively, we can 
calculate ${\pi} ^{IJ}$ as a function of 
$( {\te} \, , \, u \, , \, v \, , \, {\ba} \, , \b^{\pm})$ 
and their time derivatives $( {\dte} \, , \, {\du} \, 
, \, {\dv} \, , \, {\dba} \, , {\dbb} \, , \, {\dbc} )$ 
and then impose the constraints  ${\H} _I$. Once the constraints 
are imposed we can calculate the pre-symplectic form and determine
the canonically conjugate momenta for the diffeomorphism invariant variables.
We will choose the second alternative, because it is simpler and it gives an 
independent check of Kodama's results.

Our first step is to express the conjugate momenta ${\pi} ^{IJ}$ 
in terms of the new variables
$$
\eqalignno { {\pi} ^{IJ} &= - {\sqrt {\det g}}
\bigl ( g ^{IM} K _{MN} g ^{NJ} - g ^{IJ} \, g ^{MN} K _{MN} \bigl )
					\quad , &\eqnalign\momg \cr}
$$
where the extrinsic curvature $K _{IJ}$ is defined by
$$
\eqalignno { K _{IJ} &= {1\over 2{\N}} \, \bigl ( - {\dg} _{IJ} 
+ ( L _{\nen} g ) _{IJ} \bigr ) 
                      	  \quad . &\eqnalign\excur \cr}
$$
Our choice of foliation is such that ${\N} ^I = 0$, so that
$$
\eqalignno { K _{IJ} &= - {1\over 2{\N}} \, {\dg} _{IJ}
                        \quad . &\eqnalign\excurv \cr}
$$
The vector constraint {\badiff} is expressed in terms of the product  
$g _{IK} {\pi} ^{KJ}$ and from the equations {\momg} and {\excurv}
we calculate
$$
\eqalignno { g _{IK} \, {\pi} ^{KJ} &= {{\sqrt {\det g}}\over 2 {\N}} 
\Bigl ({\dg} _{IK}  g ^{KJ} - {\d} _I^J \, 
\bigl ( {\dg} _{LK} g ^{KL} \bigr ) \Bigr )
			\quad . &\eqnalign\prodpg \cr}
$$
Thus,
$$
\eqalignno { {\H} _1 &= 2  \, g _{2K} \, {\pi} ^{K3}  
\sim {\dg} _{2K} g ^{K3}	
			\quad , &\eqnalign\hone \cr
{\H} _2 &= - 2  \, g _{1K} \, {\pi} ^{K3} 
\sim {\dg} _{1K} g ^{K3}	
			\quad , &\eqnalign\htwo \cr
{\H} _3 &= 2  \, \bigl ( g _{1K} \, {\pi} ^{K2} -  
g _{2K} \, {\pi} ^{K1} \bigr )
\sim \bigl ( {\dg} _{1K} g ^{K2} - {\dg} _{2K} g ^{K1} \bigr ) 
			\quad . &\eqnalign\hthree \cr}
$$
In order to impose the vector constraint, we only have to calculate 
the components of the product ${\dg} _{IK} g ^{KJ}$ in terms of
our new variables {\newg} 
$$
\eqalignno { {\dg} _{IK} g ^{KJ} &= {\Bigl ( {M _E} ^T \, Q \, 
{\Md} {M _E} ^{-1} \, Q^{-1} 
\bigl ( {M _E} ^T \bigr )  ^{-1} \Bigr ) _I} ^J	\cr
&+ {\Bigl ( {M _E} ^T \, {\dot Q}  \, Q^{-1}
\bigl ( {M _E} ^T \bigr )  ^{-1} \Bigr ) _I} ^J
+ {\Bigl ( {\Md} ^T \bigl ( {M _E} ^T \bigr )  ^{-1} \Bigr ) _I} ^J
		\quad . &\eqnalign\productgg  \cr}
$$
We substitute {\newm} and {\gtcomp} into {\productgg} and after a 
straightforward calculation we find that $H_I =0$ is equivalent to
$$
\eqalignno { {\du} - v \, {\dte} &= 0 \quad , \quad {\dv} + u \, {\dte} = 0	
\quad , \quad \sinh^2 (2\sqrt3 \b^- ) {\dte} = 0\quad . &\eqnalign\vcm \cr}
$$

Now we insert {\newg} into {\momg}, and by taking into account {\vcm}, we 
obtain
$$
\eqalignno{ \pi^{IJ} (t) &= ({M_E^{-1})_I}^K \, P_{KL} (t) \, 
{((M_E^{-1})^T)^L}_J 	\quad , &\eqnalign\newgafterhi	\cr}
$$
where the matrix $P={\rm diag}(P_1,P_2,P_3)$ and it is given by
$$
\eqalignno{ P_{1} &= {{\sqrt {\det g}}\over {\N}} \,  
e^{- 2 \, ({\ba} + {\bb} + {\sqrt 3} \, {\bc})} \, 
( - 2 {\dba} + {\dbb} + {\sqrt 3} \, {\dbc} )  
			\quad , \cr
P_{2}  &= {{\sqrt {\det g}}\over {\N}} \, 
e^{- 2 \, ({\ba} + {\bb} - {\sqrt 3} \, {\bc})} \, 
( - 2 {\dba} + {\dbb} - {\sqrt 3} \, {\dbc} ) 
			\quad , \cr
P_{3} &=  - 2 {{\sqrt {\det g}}\over {\N}} \,
e^{- 2 \, ({\ba} - 2 {\bb})} \, ( {\dba} + {\dbb} ) 
			\quad . &\eqnalign\newp \cr}
$$
Note that in our calculation of $P$ we have used the constraints
{\vcm}, and therefore the expression {\newp} is valid for all solutions of the 
constraints. The equations {\vcm} have
two classes of solutions. One class is given by 
$$\du = \dv =\dte =0 \quad,$$
which corresponds to the sector $\b^- \ne 0,\dot\b^- \ne 0$. This is a generic sector
invariant under 
the Bianchi-VII group [\K]. In this sector the vector constraint
gives that $u$, $v$ and $\theta$ are constants and hence $P$ is automatically
diagonal. The second class of solutions is 
$$\du - v\dte = \dv + u\dte = \b^- =0\quad,$$ 
and it is less
obvious that $P$ is diagonal in this case, but it is a consequence of the 
constraints 
{\vcm}. Namely, $\dte$ appears in the off-diagonal part of $P$, but it is
multiplied by a factor $Q^{-1}_{1} - Q^{-1}_{2}$, which vanishes 
for $\b^- = 0$. This is the sector which is invariant under a  
group larger then the Bianchi-VII group [\K]. We will concentrate on the 
generic sector,
although we will give some brief comments about the enhanced symmetry sector.

The pre-symplectic structure ${\a}$ by definition {\baaction} is 
$$
\eqalignno { \a &= {\pi} ^{IJ} \, {\dif} g _{IJ}
			\quad . &\eqnalign\psympl \cr}     
$$
In order to calculate ${\a}$ in terms of the new variables
we substitute the expressions {\newgafterhi} and {\newp}
for the three metric $g _{IJ}$ and its conjugate momenta 
${\pi} ^{IJ}$ into the equation {\psympl},
and we obtain
$$
\eqalignno { \a &= {\Tr} \bigl ( P {\dif} Q \bigr ) +
2 {\Tr} \bigl ( PQ \, {\dif} M _E \, {M _E} ^{-1} \bigr )
                         \quad . &\eqnalign\psymplexplicit \cr} 
$$
The second term in {\psymplexplicit} can be calculated 
from the definition of the matrix
$M _E$ {\newm}, and we get
$$
\eqalignno { {\dif} M _E \, {M _E} ^{-1} &=   
\left ( \matrix { 
0 & {\dif} {\te}  & {\dif} u - v {\dif} {\te}   \cr
-{\dif} {\te} & 0 & {\dif} v + u {\dif} {\te}   \cr
0 & 0            &   0                       \cr} \right ) 
                        \quad . &\eqnalign\diffm \cr} 
$$
From {\newgafterhi} and {\newp}
we obtain that the components of the diagonal matrix 
${\bigl ( PQ \bigr ) ^I} _J$ are
$$
\eqalignno { {\bigl ( {P} {Q} \bigr ) ^1} _1 &= 
{{\sqrt {\det g}}\over {\N}} 
\bigl (- 2 {\dba} + {\dbb} + {\sqrt 3} \, {\dbc} \bigr ) 
                        \quad , &\eqnalign\pitgtone \cr  
{\bigl ( {P} {Q} \bigr ) ^2} _2 &=  \, {{\sqrt {\det g}}\over {\N}}
\bigl (- 2 {\dba} + {\dbb} - {\sqrt 3} \, {\dbc} \bigr )
                        \quad , &\eqnalign\pitgttwo \cr  
{\bigl ( {P} {Q} \bigr ) ^3} _3 &= - 2 \, {{\sqrt {\det g}}\over {\N}}
\bigl ( {\dba} + {\dbb} )
                         \quad . &\eqnalign\pitgtthree \cr}
$$
Since $PQ$ is a diagonal matrix, it is obvious that
$$
\eqalignno { {\Tr} \bigl ( PQ \, {\dif} M _E \, {M _E} ^{-1} \bigr )
&= 0                     \quad . &\eqnalign\secondterm \cr}
$$
Thus, the second term in {\psymplexplicit} is identically 
equal to zero and ${\a}$ is given by
$$
\eqalignno { \a &= {\Tr} \bigl ( P {\dif} Q \bigr )
                         \quad . &\eqnalign\psymplectic \cr} 
$$
From 
$$
\eqalignno { {\dif} Q_{1} &= 
e ^{2 \, ({\ba} + {\bb} + {\sqrt 3} \, {\bc})} 
2 ( {\dif}{\ba} + {\dif} {\bb} + {\sqrt 3} \, {\dif}{\bc} )
     			\quad , &\eqnalign\diffgtone \cr
{\dif} Q_{2} &=
e ^{2 \, ({\ba} + {\bb} - {\sqrt 3} \, {\bc})}
2 ( {\dif}{\ba} + {\dif}{\bb} - {\sqrt 3} \, {\dif}{\bc} )
     			\quad , &\eqnalign\diffgttwo \cr
{\dif} Q_{3} &= 
e ^{2 \, ({\ba} - 2 {\bb})}
2 ( {\dif}{\ba} - 2 \, {\dif}{\bb} ) 
     			\quad , &\eqnalign\diffgtthree \cr}
$$
and from {\newp}
we obtain
$$
\eqalignno { \a &= 12 \, {{\sqrt {\det g}}\over {\N}} 
\Bigl ( - {\dba} \dif {\ba} + {\dbb} \dif {\bb} 
+ {\dbc} \dif {\bc} \Bigr ) 
                       	\quad . &\eqnalign\psymplecticphys \cr}
$$
Since
$$
\eqalignno { \!\!\!\!\! {\pa} &= - \Bigl (12 \, {{\sqrt {\det g}}\over {\N}} 
\Bigr )	{\dba} \, , \, \,  
{\pb} = \Bigl (12 \, {{\sqrt {\det g}}\over {\N}} \Bigr ) {\dbb}
\, , \, \, 
{\pc} = \Bigl (12 \, {{\sqrt {\det g}}\over {\N}} \Bigr ) {\dbc} 
                       	\, , &\eqnalign\physmom \cr}
$$
the pre-symplectic form becomes
$$
\eqalignno { \a &= 
 p_0 {\dif} {\ba} + p_+ {\dif} {\bb} 
+ p_- {\dif} {\bc}  \quad . &\eqnalign\rsf \cr}
$$

Note that {\rsf} is the pre-symplectic form for the generic sector. 
The enhanced symmetry sector is a special case where $p_- =\b^- =0$, so that 
the pre-symplectic form becomes
$$
\eqalignno { \a &= 
 p_0 {\dif} {\ba} + p_+ {\dif} {\bb} 
  \quad . &\eqnalign\drsf \cr}
$$

Dynamics of the canonical pairs $( {\ba} \, , \, {\pa} \, , \,  {\bb} 
\, , \, {\pb} \, , \,  {\bc} \, , \, {\pc} )$ is defined
by the Hamiltonian constraint {\baham}.  To calculate the expression
for the Hamiltonian constraint in terms of the diffeomorphism invariant 
variables
we substitute {\newgafterhi} and {\newp} into {\baham} and obtain
$$
\eqalignno { {\H} _0 &= {1\over{\sqrt {\det g}}} 
\Bigl ( {\Tr} \, ( PQPQ ) - {\half} \, 
{\Tr} ^2 (PQ ) + {\Tr} \, (  S Q S Q )
- {\half} \, {\Tr} ^2 ( S Q ) \Bigr )
                       	\, . &\eqnalign\newham \cr}
$$
We notice that the Hamiltonian constraint is independent of the 
unphysical variables $( {\te} \, , \, u \, , \, v  )$. In addition, we already
have the components of the product ${\bigl ( PQ \bigr ) ^I} _J$, which are 
given by the equations {\pitgtone}, {\pitgttwo} and {\pitgtthree}, and 
$SQ$ is a diagonal matrix
$$
\eqalignno { \!\!\!\! {\bigl ( SQ \bigr ) ^1} _1&=
e ^{2 \, ({\ba} + {\bb} + {\sqrt 3} \, {\bc})}	\, , \, \,   
{\bigl ( S Q \bigr ) ^2} _2 =  
e ^{2 \, ({\ba} + {\bb} - {\sqrt 3} \, {\bc})}	\, , \, \, 
{\bigl ( S Q \bigr ) ^3} _3 = 0
				\, . &\eqnalign\sgt \cr}
$$
From this we obtain that the Hamiltonian constraint for the generic sector is
given by
$$
\eqalignno { \!\!\!\! {\H} _0 &=
{6 \, {\sqrt {\det g}}\over {\N} ^2} \, 
\bigl (- ( {\dba} ) ^2 + ( {\dbb} ) ^2 + ( {\dbc} ) ^2 \bigr ) 
+ {2 \, e ^{4 ({\ba} + {\bb})}\over{\sqrt {\det g}}} \, 
{\sh} ^2 ( 2 {\sqrt 3} {\bc} )
                       	\, , &\eqnalign\physhamiltonian \cr}
$$
or in terms of the canonical pairs {\physmom} 
$$
\eqalignno { \!\!\!\!\!\!\!{\H} _0 &= {1\over24 \, {\sqrt {\det g}}} 
\Bigl ( - ( {\pa} ) ^2 + ( {\pb} ) ^2 + ( {\pc} ) ^2 + 
48 \, e ^{4 ({\ba} + {\bb})} \, {\sh} ^2 ( 2 {\sqrt 3} {\bc} ) \Bigr )
                       	\, . &\eqnalign\physham \cr}
$$

The previous calculation applies also to the enhanced symmetry sector, 
but one must take
into account that $p_- =\b^- =0$. The Hamiltonian constraint then becomes
$$
\eqalignno { \!\!\!\!\!\!\!{\H} _0 &= {1\over24 \, {\sqrt {\det g}}} 
\Bigl ( - ( {\pa} ) ^2 + ( {\pb} ) ^2  \Bigr )
                       	\quad, &\eqnalign\dphysham \cr}
$$
which corresponds to the dynamics of a two-dimensional relativistic free 
particle.
 
{\bf \chapter {Diffeomorphism invariant phase space}}

Note that the reduced pre-symplectic form {\rsf} implies that the 
diffeomorphism
invariant phase space is given by $(p_\mu , \b^\mu )$ canonical pairs. However,
this is not correct, since it has been shown recently that the diffeomorphism
invariant subspace is larger, because it contains the moduli parameters, which
are associated with the global degrees of freedom of the metric 
[\J, \K]. This is a consequence of the fact
that Bianchi metrics on compact spatial slices represent compact Reimannian 
manifolds
with non-trivial topology, and such manifolds 
can be locally diffeomorphic, but not globally diffeomorphic.

As a result one cannot define globally a Bianchi metric, so that one needs a 
notion of locally homogeneous spacetime [\AS,\JA].
A locally homogeneous Bianchi spacetime $M$ with a symmetry 
group $G$ can be represented as $\tilde M / K$, where $\tilde M$ is a 
homogeneous
Bianchi space-time with a simply connected spatial section, and $K$ is a 
discrete
subgroup of $G$ [\J]. In the following we will review Kodama's results 
[\K], since his approach is suitable for the Hamiltonian formalism.

The symmetry group $\tilde G$ of the covering space 
$\tilde M$ is defined as
$$\tilde G = \{ f \in Diff(\tilde M) \mid f_* \tilde\Phi = \tilde\Phi \}\,, $$
such that it is isomorphic to $G$, or it is the smallest possible group which 
contains $G$, and it acts transitively on $\tilde M$.
$\tilde\Phi = j^* \Phi$ is the pullback of the canonical data $\Phi$ on $M$ and
$j: \tilde M \to M$ is the covering map. 

The moduli space can be parametrized with finitely many parameters 
$\{ \l^a \}$, which are determined from the covering maps 
$j_\l : \tilde M \to M$ such that $ j^*_\l (\pi_1 (M)) = K_\l$ is isomorphic 
to $K$ and it is a subgroup of $\tilde G$, where $\pi_1 (M)$ is the fundamental
group of $M$.  
One then has to find conjugacy classes
of $K_\l$ under the
homogeneity preserving diffeomorphisms HPDG
$$ HPDG = \{ \tilde f \in Diff(\tilde M) \mid \tilde f \tilde G \tilde 
f^{-1} = \tilde G\}\quad.$$
This procedure gives that $K_\l = f_\l K_0 f_\l^{-1}$, where $K_0$ is a 
reference point,
such that $\tilde M / K_0$ is identified with $M$, and $f_\l$ is a linear transformation
on $\tilde M$ such that
$$
\eqalignno { f_\l^* \chi^I &= F^I_J \chi^J \cr  f_\l^* g_{IJ} &= (F^T)_I^K g_{KL} F_L^K 
\cr f_\l^* \pi^{IJ}  &= (F^{-1})^I_K \pi^{KL} ((F^{-1})^T)_L^J \cr 
f_\l^* \sqrt{\det g} &= (\det F)(\det\chi)\sqrt{\det g_{IJ}} 
                       	\, . &\eqnalign\mot \cr}$$
From {\mot} one obtains that the pre-symplectic form {\psymplexplicit} becomes
$$ \a = {\Omega(\l)\over V}(\pi^{IJ} dg_{IJ} + 2 C_a (\l) \pi^{IJ} g_{IJ} 
d\l^a ) \quad,$$
where $V=\int_{D_0} d^3 x \det \chi$,
$$ \Omega (\l) = \int_{D_0} d^3 x (\det\chi) (\det F) \quad,\quad
C_a (\l) ={1\over\Omega} \int_{D_0} d^3 x (\det\chi)(\det F)\partial_{a} F F^{-1}
\quad, $$
and $D_0$ is the fundamental region of the action of $K_0$ on $\tilde M$.
The Hamiltonian constraint is rescaled by $V/ \O(\l)$.

In the case of Bianchi VII model, $G = VII^+_0$, where $+$ fixes the orientation of 
the spatial section. $M$ is a three-torus $T^3$, which can be 
represented as $E^3 /K$ where $E^3 $ is the Euclidian space, and 
$K= {\bf Z}^3$.
$\tilde G = G \tilde\times D_2$, where $D_2$ is the dihedral group
and $\tilde\times$ denotes a semi-direct product. $K_\l$ are
represented by $GL(3,{\bf R})$ matrices, whose last row is given by $(l\pi,m\pi,n\pi)$,
$l,m,n \in {\bf Z}$. By using HPDG and modular transformations, $K_\l$ can be put into
form
$$\eqalignno { 
K_\l &=\left ( \matrix { 
X & Y  & Z   \cr
0 & X^{-1} & W   \cr
0 & 0            &   n\pi  \cr} \right )
\quad,\quad X>0\quad,\quad n\ge 0 \quad. &\eqnalign\modp \cr}$$ 
Therefore $X,Y,Z,W$ and $n$ represent the modular parameters. 
The $K_0$ is associated with $X = 1$, $Y=Z=W=0$, and the deformation map $f_\l$ is
given by the matrix
$$\eqalignno {
A &=\left ( \matrix { 
X & Y  & Z   \cr
0 & X^{-1} & W   \cr
0 & 0            &   1  \cr} \right )
\quad. &\eqnalign\defm \cr}$$
The fundamental region is $D_0 =\{ 0\leq x,y \leq 1 , 
0\leq z \leq n\pi \}$.

From {\defm} one obtains
$$\eqalignno{ F  &= R_3 (z) A R_3 (-z)
\quad, &\eqnalign\bsf \cr}$$
where $R_3$ is the rotational matrix around $z$-axes. 
The correction in the pre-symplectic form due to modular parameters is given by
$$\eqalignno{ Tr ( \dot F F^{-1} p g)  =& \Bigl[ {\dot X\over X}
 \cos(2z) - 
\half (X\dot Y - Y\dot X)\sin (2z)
\Bigr] (Q_{1} P_{1} - Q_{2} P_{2})\, . &\eqnalign\bsfa \cr} $$
However, when {\bsfa} is integrated over $D_0$, one obtains zero, and hence 
$\d\a$ vanishes [\K].
Therefore the moduli parameters do not enter into the reduced pre-symplectic 
form, and hence they 
will not influence the dynamics of the local degrees of freedom.
 
{\bf \chapter {Hamiltonian Constraint}}

The action for the generic sector of the Bianchi VII$_0$ model can now be 
expressed in terms of the
canonical variables $(p _{\mu}, {\b}^{\mu} )$ as  
$$
\eqalignno { S &= \int _{t_0}^t {\dif} t \, \Bigl ( p _{\mu}
{\db} ^{\mu} - {\tilde N} \, {\tilde H}_0  \Bigr ) 
			\quad , &\eqnalign\bviiaction \cr}
$$
were  ${\tilde N}$ is the Lagrange multiplier and 
${\tilde H}_0 = \sqrt{\det g}H_0 /4$ is the
rescaled Hamiltonian
constraint {\physham}. This system is reparametrization invariant
{\ie} it has a symmetry generated by the constraint ${\tilde H}_0$
$$
\eqalignno { \d p &= \e \, \{ {\tilde H} _0 , p \}  \, ,  
\quad \d q = \e \, \{ {\tilde H} _0 , q \}  ,  \quad \d {\tilde N} = 
{{d\tilde N}\over dt}
			\quad , &\eqnalign\simtransf \cr}
$$
where $\e$ is the parameter of the transformation.

In order to find the dynamics of the physical degrees of freedom, we need to
fix the reparametrization gauge symmetry {\simtransf}. As discussed in 
[\MM], this
type of gauge-fixing requires the specification of the time variable, in 
addition to the usual requirement that the Faddeev-Popov determinant is 
non-zero. In the case of our system, the analysis simplifies if we introduce
new canonical coordinates
$$
\eqalignno { T &= 4 \, ({\ba} + {\bb})- \ln 16	\, , \quad
p _T = {1\over 8} \, \, ( p _0 + p _+ ) \, , \cr
q^1 &= 6 ({\ba} - {\bb})	\, , \quad
p_1 = {1\over 12} \, ( p _0 - p _+ ) \, , \cr
q  &= 4 \sqrt 3  \, {\bc}	\, , \quad
p  = {1\over 4 \sqrt 3} \, p _- \, . &\eqnalign\catrans \cr}
$$ 
The Hamiltonian constraint now becomes
$$
\eqalignno { {\tilde H} _0 &= - p _T \, p _1 + 
\half p^2 + 8 e^T \, \sh^2 \bigl({q\over 2}\bigr) \, , 
&\eqnalign\hconst \cr}
$$
Since {\hconst} is independent of $q^1$, it follows that
$p_1$ is a constant $c$. On the other hand, the Lagrange multiplier 
$\tilde N$ plays the 
role of the one-dimensional metric on the world-line parametrized by $t$, and 
hence it can be set to a positive constant via {\simtransf}. Since the equation
for $T$ is given by
$$
\eqalignno { \dot{T} + c \tilde N &= 0
			\quad, &\eqnalign\teq \cr}
$$ 
then one can choose the following gauge
$$
\eqalignno { T &= t \quad,\quad \tilde N =-1/c \quad. &\eqnalign\gc \cr}
$$ 
Note that the gauge choice {\gc} requires $p_1 = c < 0$, in order for 
$\tilde N$ to be positive and finite. Configurations with $p_1 = -c$
are physically equivalent to $p_1 = c$ configurations, because 
our system is invariant under time-reversal (configurations
with $p_1 = -c$ have $T = -t$). The configurations with $p_1 = 0$ are 
excluded, since they
belong to the enhanced-symmetry sector $p_- = \beta^- =0$.

The equation {\gc} defines the required gauge choice, and therefore the
Hamiltonian of the physical degrees of freedom is given by solving the
Hamiltonian constraint for $p_T$
$$
\eqalignno { p _T &={H^* \over c} = {1\over c } \, \Bigl (  
\half p^2 + 8e^{t} \, \sh^2 \bigl ( {q\over 2} \bigr) 
\Bigr )	\, . &\eqnalign\pt \cr}
$$
Hence the physical phase space is given by the $(p, q)$ canonical pairs,
and the corresponding Hamiltonian is given by $H^*$. The dynamical consistency
of the gauge choice {\gc} can be checked explicitly, by comparing the 
equations of motion for $p_T , p , q$ from {\hconst} with the 
corresponding equations coming from the reduced Hamiltonian $H^*$. For example
$$
\eqalignno { \dot{p_T} &= -8\tilde N e^T \sh^2 \bigl ( {q \over 2} \bigr ) 
 ={\partial p_T \over \partial t}
+ \{ H^* , p_T \}^* = {\partial p_T \over \partial t} \quad , 
&\eqnalign\dc \cr}$$  
where $\{,\}^*$ is the Poisson bracket with respect to the reduced phase space
variables  $(p, q)$.
 
From $H^*$ we get $p = \dot{q}$ and $\dot p = - 4 e^{t} \sh q $, so that 
$$
\eqalignno { {d^2 q \over dt^2 } + 4 e^{t} \sh q  &= 0
			\quad . &\eqnalign\pp \cr}
$$
This is a Painlev\'e III equation [\IN]. One can put {\pp} into the
standard form via time redefinition $\t = e^t$
$$
\eqalignno { {d\over d\t}\bigl( \t {d q \over d\t }\bigr) + 4 \sh q  &= 0
			\quad . &\eqnalign\piii \cr}
$$

Note that the standard form {\piii} could have been also obtained 
directly by choosing the gauge $T = \log t$, $p_1 = -c$ and  
$\tilde N = -1/(ct)$. 

{\bf \chapter {Zero-curvature representation of Painlev\'e III}}

By using a more general framework of Belinskii-Zakharov inverse scattering 
method 
for the spacetimes admitting two commuting spacelike Killing vectors [\BZ],
Belinskii and Francaviglia showed that the Einstein equations for certain 
Bianchi models admit a zero-curvature representation [\BF]. 
In this section we will review their approach for Bianchi I, II, VI$_0$ and VII$_0$ 
models. A particular attention will be given to the  Bianchi VII$_0$ model.
We will show that within the framework of the inverse scattering method,
the dynamics of the Bianchi VII$_0$ model is given by the
Painlev\'e III equation {\piii}. This is in complete accordance with the 
results
from the previous section. In addition, we will derive a zero-curvature 
representation 
for this Painlev\'e III equation.

We begin with a brief discussion of Belinskii and Zakharov method for
the midi-superspace models that are characterized by the existence of a 
two-parameter Abelian group of motions with two spacelike Killing vectors [\BZ]. 
Let us choose coordinates adapted to the action of the symmetry group so 
that the metric assumes the following form 
$$
\eqalignno{ ds ^2 &=  - f \ dt ^2 + f \ dz ^2 + g _{ab} \ dx ^a dx ^b \ ,
&\eqnalign\dvakv	\cr}
$$
where $a,b = 1,2$, $\{ x ^0, x ^1, x ^2, x ^3 \} =\{ t, x, y, z \}$,
$f$ is a positive function and $g _{ab}$ is a symmetric two-by-two matrix.
The function $f$ and the matrix $g _{ab}$ depend only on the co-ordinates
$\{ t,z \}$, or equivalently on the null co-ordinates $\{ {\x} , {\y} \} =
\{ \half ( z + t ), \half ( z - t ) \}$.
There is a freedom to perform the co-ordinate transformations
$$
\eqalignno{ \{{\x} , {\y} \} &\to \{ \tilde\xi(\xi),\tilde\eta(\eta) \}
\, . &\eqnalign\conf \cr}
$$
It is easy to see that the transformations {\conf} preserve both
the conformally flat two-metric $f ( - dt ^2 + dz ^2)$ and the positivity of
the function $f$ if $\pde {\tilde\xi} {\x} \, \pde {\tilde\eta} {\y} > 0$.

The complete set of vacuum Einstein equations for the metric
{\dvakv} decomposes into two groups of equations [\BZ]. The first group
determines the matrix $g _{ab}$ and can be written as a single
matrix equation
$$
\eqalignno{ \pde { \bigl ( \a \, \pde g {\x} \, {g ^{-1}} \bigr ) } {\y}
+ \pde { \bigl ( \a \, \pde g {\y} \, {g ^{-1}} \bigr ) } {\x} &= 0
\, ,  &\eqnalign\sigmamodel \cr}
$$
where ${\a } ^2 = \det g$ and $\{ {\x} , {\y} \}$ are the null co-ordinates.
The second group of equations determines the function $f( \x , \y )$
in terms of a given solution of {\sigmamodel}:
$$
\eqalignno{
\pde {(\ln f)} {\x} &= { \partial ^2 _{\x} {(\ln {\a})}\over{ \partial _{\x}
{(\ln {\a})} }} + { 1\over{ 4 {\a} \, {\a} _{\x} }} \tr A ^2 
\, , & \eqnalign\fa \cr
\pde {(\ln f)} {\y} &= { \partial ^2 _{\y} {(\ln {\a})}\over{ \partial _{\y}
{(\ln {\a})} }} + { 1\over{ 4 {\a} \, {\a} _{\y} }} \tr B ^2 
\, , & \eqnalign\fb \cr}
$$
where ${\a} _{\x} = \pde {\a} {\x}$, ${\a} _{\y} = \pde {\a} {\y}$ and
the matrices $A$ and $B$ are defined by
$$
\eqalignno{ A = - {\a} \ \pde g {\x} \ g ^{-1},
\quad & \quad B = {\a} \ \pde g {\y} \ g ^{-1} \, . & \eqnalign\ab \cr}
$$
The dynamics of the system is thus essentially determined by 
the equation {\sigmamodel}. 
By taking the trace of the equation {\sigmamodel} and by using the 
definition for ${\a}$, we obtain
$$
\eqalignno {  {\a} _{\x \y} &= 0   \, . &\eqnalign\wave \cr}
$$
The two independent solutions of this equation are
$$
\eqalignno{ {\a} &= c( {\x} ) + d ( {\y} ) 
\ , \quad   {\b}  = c( {\x} ) - d ( {\y} ) \, . 
& \eqnalign\wavesol  \cr}
$$
By using the transformations {\conf}, one can bring the functions $c({\x})$
and $d({\y})$ to a prescribed form. However, we will consider the general
form without specifying the functions $c({\x})$ and $d({\y})$ in advance.

The crucial step in the inverse scattering method
is to define the linearized system whose integrability conditions
are the equations of interest, in our case the equation {\sigmamodel}. 
Following ref. [\BZ], we define the two differential operators
$$
\eqalignno{ D _1 &= \pde {\z} {\x} - {{2{\a} _{\x} \, {\l}}\over{ {\l} - {\a} }} 
\, \pde {\z} {\l} ,  & \eqnalign\da \cr
D _2 &= \pde {\z} {\y} + {{2{\a} _{\y} \, {\l}}\over{ {\l} + {\a} }} 
\, \pde {\z} {\l}, & \eqnalign\db \cr}
$$
where ${\l}$ is a complex parameter independent of the co-ordinates
$\{ {\x} , {\y} \}$. It is straightforward to see that the
differential operators $D _1$ and $D _2$ commute since ${\a}$
satisfies the wave equation {\wave}
$$
\eqalignno{ [ D _1 , D _2 ] &= {\a} _{\x \y} \
{(2 \l) ^2\over{\l ^2 - \a ^2}} \ \partial _{\l} = 0
\ .  & \eqnalign\com \cr}
$$
The next step is to consider the following linear system
$$
\eqalignno{ D _1 {\f} &= {A\over{ {\l} - {\a} }} {\f}
\, ,  & \eqnalign\ls \cr
	    D _2 {\f} &= {B\over{ {\l} + {\a} }} {\f}		
\, ,  & \eqnalign\lsb \cr}
$$
where ${\f} ( {\l} , {\x} , {\y} )$ is a complex matrix function,
and the real matrices $A$, $B$ and the real function ${\a}$
do not depend on the complex parameter ${\l}$. The integrability
conditions for the system {\ls} and {\lsb} are given by the equation 
{\sigmamodel}. 
Furthermore, a solution  
${\f} ( {\l} , {\x} , {\y} )$ yields a matrix $g ( \x , \y )$ 
that satisfies the original equation
{\sigmamodel}. Namely, the matrix $g ( \x , \y )$ is given by 
$$
\eqalignno { g ( \x , \y ) &= {\f} ( {\l} , {\x} , {\y} ){\big |} _{{\l}= 0} 
\ .  & \eqnalign\gpsi \cr}          
$$
In order to take into account that $g ( \x , \y )$ is real and symmetric 
we have to impose two additional conditions, see [\BZ]. Also, it is easy 
to see that the equations {\ls} and {\lsb} for ${\l}= 0$, imply 
equations {\ab}.

Although Belinskii and Francaviglia formulation is more general [\BF], we will
discuss only type {\bf A} Bianchi models which are compatible with the inverse 
scattering method. It is not difficult to show that Bianchi types I, II, 
VI$_0$ and VII$_0$ admit the representation {\dvakv}. Namely, the metric for 
these Bianchi types has the form
$$
\eqalignno { ds^2 &= - dT^2 + g_{ij} \ dx^i dx^j \ , 
& \eqnalign\bma \cr}
$$
where
$$
\eqalignno { g_{ij} &= g _{IJ} \  {{\vf} ^I} _i \ {{\vf} ^J} _j \ . 
& \eqnalign\bmb \cr}
$$
For these models it is always possible to have the one forms ${\vf} ^I$ in the following
form 
$$
\eqalignno { {\vf} ^1 &= {l ^1} _1 \ {\dif} x + {l ^1} _2 \ {\dif} y \ , \quad 
             {\vf} ^2  = {l ^2} _1 \ {\dif} x + {l ^2} _2 \ {\dif} y \ , \quad
             {\vf} ^3  = {\dif} z  \ ,  & \eqnalign\baforms \cr}
$$
where ${l ^a} _b$ are functions of $z$ only. Let us consider the two-by-two
matrix
$$
\eqalignno {   l &=  \pmatrix { {l ^1} _1 & {l ^1} _2 \cr
				{l ^2} _1 & {l ^2} _2 \cr}
\quad .  &\eqnalign\matrixl \cr}
$$
An important consequence of the  Maurer-Cartan equations for the one forms 
${\vf} ^I$
 is that the matrix $l$ satisfies the following linear differential equation
$$
\eqalignno { {dl\over dz} &= C^T \e l \quad, &\eqnalign\bsa \cr}
$$
where the matrix $C$ is the same matrix as the upper two-by-two block on the 
principal
diagonal of the matrix $S ^{IJ}$ defined in the equation {\stconst}. $\e$ is 
the antisymmetric matrix with $\e_{12} = 1$.

After a time redefinition $t = t (T)$, the metric {\bma} can be written 
in the form 
$$
\eqalignno { ds^2 &= f(t) \ (-dt^2 + dz^2 ) + g_{ab}(t,z) \ dx^a dx^b \quad. 
&\eqnalign\csm \cr}
$$
Here $f$ is a function of $t$ only, and
$$
\eqalignno { g ( t , z ) &= l ^T (z) \g (t) l (z) \quad, &\eqnalign\hr \cr}
$$
where $l$ is given by {\matrixl} and $\g$ is a two-by-two symmetric matrix. 
Notice that now
$$
\eqalignno { \a ^2 &= ( \det l ) ^2 \det \g \ . &\eqnalign\al \cr}
$$
Moreover, for these models, the determinant of the matrix $l$ is always
equal to one, {\ie} $\det l =1$, so that
$$
\eqalignno { \a ^2 ( t )&= \det \g ( t) \ . &\eqnalign\al \cr}
$$
In addition, $\a$ has to satisfy the equation {\wave}, which now reads
$$
\eqalignno { \ddot {\a } ( t )&= 0 \ . &\eqnalign\alz \cr}
$$
Hence, $\a$ can only be a linear function of time.  

As Belinskii and Francaviglia have showed [\BF], the linearized system {\ls} and {\lsb}
can be simplified for the models described by the metric {\csm}. The first step
is to define a two-by-two matrix function $\varphi$ by
$$
\eqalignno { \psi &= l^T \, \varphi \ l \ , & \eqnalign\hphi \cr}
$$
and a constant two-by-two matrix
$$
\eqalignno { R &= \e \ C \ . & \eqnalign\hR \cr}
$$
The second step is to substitute {\hr} into {\ab} and use the definition
of the coordinates $\x$ and $\y$. Then the results of these calculations,
together 
with the definition {\hphi}, can be used to simplify the equations {\ls} and 
{\lsb}.
The crucial step in which a simplification occurs is the coordinate
transformation $\{ t , z , \l \} \to \{ t , w , \l \}$, where $w$ is given by
$$
\eqalignno { w &= {\half} \bigl ( {\a ^2\over \l} + 2 \b 
+ \l \bigr ) \ . & \eqnalign\hw \cr}
$$
To perform this co-ordinate transformation we can use $\a$ and $\b$
as given by {\wavesol}. The linear system after this co-ordinate transformation 
involves only derivatives in $t$ and $\l$ since all the terms involving 
derivatives in $w$ are canceled. Finally, it is useful to make some simple 
linear combinations of the two equations and to use the fact that 
$\a$ is a linear function of time. In this way one obtains a new linear system
$$
\eqalignno { \pde \varphi t  &= {\a\over \l} \ \bigl ( \g R ^T \g ^{-1} \varphi
- \varphi R ^T	\bigr ) \quad , \cr 
\pde \varphi \l &=  {1\over 2{\dot \a}} \ \bigl ( - R \varphi - \varphi R ^T
+ {\a\over \l} {\dot \g} \g ^{-1} \varphi + {\a ^2\over \l ^2} \varphi R ^T
- {\a ^2\over \l ^2} \g R ^T \g ^{-1} \varphi \bigr )
\quad . &\eqnalign\lsh \cr}
$$
Although the matrix function $\varphi ( t , \l , w)$ depends 
on all 
three variables, the right-hand side of the system {\lsh} does not have any 
$w$ dependence.  

The integrability condition for the system {\lsh} is 
$$
\eqalignno { {1\over \a}{d\over dt}\bigl( \a \dot\g \g^{-1}\bigr)  &= 
R\g R^T \g^{-1}
- \g R^T \g^{-1} R \quad . &\eqnalign\zcb \cr}
$$
To derive the equation {\zcb} from the system {\lsh} it is necessary to use
the fact that for these models $\a$ is a linear function of time.

Equivalently, one can derive the equation {\zcb} by a direct substitution of 
the formula
{\hr} into  equation {\sigmamodel}. A straightforward calculation, using the 
definition 
of $\x$, $\y$, the equation {\bsa} and the fact that $\a$ is a function of 
time 
only, yields the equation {\zcb}. Thus the dynamics of the these Bianchi 
models is essentially 
determined by the equation {\zcb}. Furthermore we have confirmed that the 
linear system 
{\lsh} corresponds to the Bianchi models under consideration.

Let us now consider Bianchi VII$_0$ model. In that case the spatial 
hyper-surface is a three torus $T ^3$. As we have shown,
the modular parameters do not affect the dynamics of the local 
degrees of freedom, and hence the equations {\zcb} and {\lsh}
will be correct dynamical equations. 

The matrix $l$ for Bianchi VII$_0$ model is  given by
$$
\eqalignno {   l &=  \pmatrix { \cos z & \sin z \cr
			      - \sin z & \cos z \cr}
\ ,  &\eqnalign\matrixlbs \cr}
$$
and therefore $R$ is 
$$
\eqalignno {   R &=  \pmatrix { 0 & 1 \cr
			      - 1 & 0 \cr}
\ .  &\eqnalign\matrixrbs \cr}
$$
We can take $\g$ to be diagonal
$$
\eqalignno { \g &= \pmatrix { a ^2 & 0    \cr
		              0    & b ^2 \cr}
\quad .  &\eqnalign\matrixab \cr}
$$
We also choose $\a = t$, thus the following relation 
between the functions $a$ and $b$
$$
\eqalignno { \a &= a \, b = t \ .  &\eqnalign\abt \cr}
$$
In order to derive the differential equation which defines the dynamics
for this model we substitute formulas {\matrixrbs} and {\matrixab}
into equation {\zcb} and use the relation {\abt} in order to eliminate
the function $b$. A straightforward calculation yields the following
scalar equation
$$
\eqalignno { {1\over t}{d\over dt}\bigl( 2t \dot a /a \bigr)  &= 
t^2  a^{-4}
- a^4 t^{-2} \ . &\eqnalign\zcp \cr}
$$
The redefinitions $\t = t^2/4$ and $e^q = a^4 /4\t$ then give the Painlev\'e III
equation {\piii}. If we define $f = e^q$ then the equation {\piii} becomes
$$
\eqalignno { {d ^2 f\over d \t ^2} &= {1\over f} \Bigl ( {d f\over d \t} 
\Bigr ) ^2 - {1\over \t} \Bigl ( {d f\over d \t} \Bigr ) + {2\over \t}
 \Bigl ( -f ^2 + 1 \Bigr ) \ . &\eqnalign\piiic \cr}
$$
The equation {\piiic} is the canonical form of the  Painlev\'e III equation with the
coefficients $\a = -2$, $\b = 2$, $\g = \d = 0$, see [\IN].  

{\bf \chapter {Conclusions}}

The dynamics of the generic sector of the Bianchi VII$_0$ model is given by 
a Painlev\'e III equation, and therefore it is an integrable model. The 
moduli parameters do not affect the dynamics of this sector, and hence do
not spoil the integrability. Therefore the original claim by Belinskii and 
Francaviglia that the Bianchi VII$_0$ model is integrable is shown to be 
correct. Furthermore, by using their results, we have found a
zero-curvature representation of the corresponding Painlev\'e III equation.  
 
In the enhanced symmetry sector we have obtained
a linear dynamical equation. We expect that the moduli parameters
would not spoil this, although a thorough investigation of this point would be
necessary.

The result that a Painlev\'e III equation appears as the dynamical equation
of the local degrees of freedom in the Bianchi VII$_0$ model can be used to 
obtain
information both about the Painlev\'e III equation and about the physical
properties of the model. 

As far as the theory of Painlev\'e III equation is considered, 
the linear system {\lsh} represents a new tool for the
study of Painlev\'e III equation. This linear system is different from the 
linear system which is
used for the study of Painlev\'e III equation within the isomonodromic 
deformation method
[\FN,\JMU,\I]. However, it remains to be explored what are the advantages of 
the new linear system.   

On the cosmology side, one can now examine the physical properties of the 
solutions, 
like small and large time asymptotic, as well as the singularities, 
since these properties 
of the Painlev\'e III solutions
have been thoroughly studied [\I]. In addition, the
quantization of this model should be straightforward, since
the reduced phase space Hamiltonian can be promoted into a Hermitian operator.
However, it is not clear whether one can find exact solutions for the quantum 
dynamics, since the Hamiltonians at different times  do not commute. 


We would like to thank Guillermo Mena Marug\'an for discussions and comments.
N.M. was partially supported by  grants PBIC/C/MAT/2150/95 and \hfil\break
PRAXIS/2/2.1/FIS/286/94. A.M. was supported by the grant 
PRAXIS \hfil\break  XXI/BCC/18981/98 from the Portugese Foundation for
 Science and 
Technology.

\refout

\bye